\documentclass[pra,onecolumn, showpacs, showkeys, secnumarabic, aps, amsmath, amssymb, nofootinbib, superscriptaddress, longbibliography, floatfix, table-of-contents, dblfloatfix]{revtex4-2}

\usepackage[utf8]{inputenc}
\usepackage[pdftex]{graphicx}
\usepackage{mathrsfs}
\usepackage[colorlinks, breaklinks, urlcolor={blue}, linkcolor={blue}, citecolor={blue}]{hyperref}
\usepackage{array}
\usepackage{amsmath}
\usepackage{type1cm}
\usepackage[english]{babel}
\usepackage{lmodern}
\usepackage{microtype}
\usepackage{booktabs}
\usepackage{caption}
\usepackage{braket}
\usepackage{xcolor}
\usepackage{orcidlink}

\begin{document}
\title{Resilience of Quantum Teleportation Fidelity for Bipartite Mixed States near Schwarzschild and Dilaton Black Holes}

\author{Abhijit Mandal\orcidlink{0000-0001-7101-9495}}
\email[]{a.mandal1.tmsl@ticollege.org}
\affiliation{Department of Mathematics, Techno Main Salt Lake (Engg. Colg.), \\Techno India Group, EM 4/1, Sector V, Salt Lake, Kolkata  700091, India}

\author{Sovik Roy \orcidlink{0000-0003-4334-341X} }
\email[]{s.roy2.tmsl@ticollege.org}
\affiliation{Department of Mathematics, Techno Main Salt Lake (Engg. Colg.), \\Techno India Group, EM 4/1, Sector V, Salt Lake, Kolkata  700091, India}

\begin{abstract}
\noindent 
We investigate the robustness of quantum teleportation in the presence of strong gravitational fields by analyzing bipartite mixed states derived from tripartite GHZ and W-class states near black hole event horizons. Considering a scenario where two observers approach the horizon of either a Schwarzschild or a Garfinkle–Horowitz–Strominger (GHS) Dilaton black hole while a third remains in flat space, we quantify the teleportation fidelity of the resulting bipartite channels after tracing out one party. Through the quantization of Dirac fields and Bogoliubov transformations, we compute the teleportation fidelity under the influence of Hawking radiation and spacetime curvature. Our results show that while entanglement degrades, teleportation fidelity remains above the classical threshold of $f>\frac{2}{3}$ for channels derived from W-class states, but not for GHZ-derived states. This indicates that quantum teleportation can remain feasible near black holes provided the initial entangled state retains useful bipartite entanglement.
\end{abstract}


\maketitle
\section{Introduction:}\label{sec:introduction}
\noindent Ever since the phoenix like rise of quantum mechanics after John Bell's seminal paper \cite{bell1964}, scientists all over the world were trying to understand the \textit{spooky action at a distance} and at the same time implementing the non-local feature of quantum states in various branches of physics, one such branch being information processing science. Entanglement is a non-local feature which can be used as a resource and the success of quantum information protocols relies on our ability to preserve it \cite{horos2009}.  Many quantum information protocols like teleportation \cite{bennett1993}, dense coding \cite{bennett1992},quantum key distribution \cite{bennett1984} and many more \cite{nielsen2010} thus emerged and were extensively studied throughout the past years, where entanglement played a decisive role. Initially, bipartite pure and mixed quantum states were used as information processing channels \cite{roy2010}. Seeing the success trend of bipartite systems, now a days, study of multi-partite quantum states have got tremendous thrust \cite{az2024,gw2024,wu2023}. Study of entanglement of different multi-partite states under varieties of physical conditions is therefore very essential. Among them, tripartite states, also play significant roles in quantum information science \cite{cradha2024,anu2024,abhi2024,royradha2024}. In tripartite scenario, there are two in-equivalent class of states viz. $GHZ$ and $W$ class. Based on Stochastic Local Operations and Classical Communications (SLOCC), $GHZ$ and $W$ are two in-equivalent classes of tripartite entangled states in the sense that one cannot be converted into the other \cite{ghzstate,wstate,coffman2000}. A $GHZ$ state is genuinely tripartite entangled state and hence when any one of its qubits is lost, the resulting state completely loses its entanglement. On the other hand, in $W$ state, genuine tripartite entanglement is missing whereas the entanglement is distributed in a bipartite way. $GHZ$ class and $W$ class of states will be defined in the later sections. This genuine tripartite entanglement is not often responsible for whether the state can be used as essential quantum teleportation channel. This means that, as $GHZ$ state does not hold bipartite entanglement in it while $W$ state does, the genuine tripartite entanglement being missing in $W$ state does not imply that its bipartite counterpart can't be used as quantum teleportation channel. Rather the case is opposite, which was shown in several works \cite{roy2010,roy2023,roy2024}. Let's explore the intriguing behavior of bipartite mixed states, derived from tripartite pure states, within the gravitational landscapes of Schwarzschild and Dilaton black holes. Our investigation aims to determine if these derived states retain their efficacy as quantum channels for reliable quantum teleportation, even amidst such extreme astrophysical environments. Black holes are perhaps the most tempting objects of Einstein’s General Relativity. For spacetime containing black holes one must encounter a spacetime singularity. This singularity, which is known as black hole singularity, may be regarded as the place where all the known laws of physics developed on a classical background break down. In a spacetime containing black hole, there exists a boundary known as the event horizon, which acts as a trapped surface such that all future directed null geodesics orthogonal to it are converging. General relativity initially predicted the existence of black holes, and subsequent astronomical observations have indirectly confirmed their presence. Despite significant progress in the study of black holes, many of their mysteries remain unsolved. From a classical perspective, when a particle crosses the event horizon of a black hole, it is trapped, making escape impossible. However, when quantum effects are considered, particles within the black hole are predicted to gradually escape, leading to what is known as Hawking radiation \cite{hawking1972}. This phenomenon serves as a crucial link between quantum mechanics and gravity and lies at the core of the black hole information paradox \cite{Hawking:1975vcx,Hawking:1976ra,Bombelli:1986rw}. In particular, quantum entanglement influenced by Hawking radiation may play a key role in resolving the information paradox of black holes. The intersection of black holes and quantum entanglement is a frontier of modern physics and has received considerable attention \cite{Fuentes-Schuller:2004iaz,Pan:2008yr, Adesso:2007wi, Alsing:2006cj, mm2010} from researchers in recent years. The existence of black holes were often debated until very recently its presence has been detected by ``Event horizon telescope collaboration'' \cite{ka1,ka2,ka3}. We in this paper have thought of a theoretical model where Alice ($A$), Bob ($B$) and Cliff ($C$), traveling in a space-ship, suddenly come in contact with a nearby black hole (often thought of in the spectrum of movies). They share among themselves well-known tripartite states like $GHZ$ or $W$ where each of the three qubits are respectively held by them. We assume that, while wandering, Bob ($B$) and Cliff ($C$) come near the event horizon, while Alice ($A$) still remains in the flat region.. The question is, will Bob (or Cliff) be able to use the tripartite channel to share quantum information to Alice? The one way is to trace out qubits from the channel, which either of the parties Bob ($B$) (or Cliff ($C$))  holds with Alice ($A$), and then the third party i.e. either Cliff ($C$) (or Bob $B$) that remains entangled can then use the bipartite mixed entangled channel with Alice ($A$) to share quantum information. We have investigated this feature here in this paper. Here, our work is significant in two ways. (a) We have considered two black holes viz. Schwarzschild Black Hole and Dilaton black hole; (b) the main motivation of the study is to investigate the derived bipartite mixed state, from tripartite classes of pure states that have close encounter with the regions surrounding the said black holes, from the perspective of the states' competence as efficient quantum teleportation channels. This effectiveness of the channels have been justified in terms of teleportation fidelity.\\\\
Prior studies on quantum teleportation process and its fidelities in black hole spacetimes \cite{gw2024,xiao2020, wu2023,Liu:2025qrw,Vakili:2026bbu} focused on bipartite pure states or tripartite systems, overlooking derived bipartite mixed states from tracing out horizon affected parties. In this work, we extend this in three ways: (i) comparing teleportation fidelity in Schwarzschild and GHS Dilaton black holes with identical tripartite states, revealing how the Dilaton parameter($D$) introduces qualitatively different degradation patterns; (ii) explicitly distinguish between prototype $W$ states and non-prototype $W_1$ states, demonstrating that their different entanglement structures lead to distinct teleportation fidelity behaviors; and (iii) showing fidelity exceeds classical thresholds for $W$ class states despite concurrence loss, over wide parameters. This understanding offers practical insights for quantum communication networks that may operate in strong gravitational fields.\\\\
The organization of this paper is as follows. In section II, we discuss the vacuum structure for Dirac fields in a Schwarzschild and GHS Dilaton spacetime. Section III discusses genuine tripartite entanglement of SLOCC class of states viz. $GHZ$ state and $W$ state and also studies the teleportation fidelities of bipartite counterparts generated from these tripartite states after tracing out parties. This is followed by conclusion in section $IV$.

\section{Methods}
\subsection{Quantization of Dirac fields in Schwarzschild and Dilaton spacetime:}
\noindent Our investigation begins with the Schwarzschild black hole and will later extend to the Dilaton black hole. The Schwarzschild black hole serves as a key model in quantum entanglement studies, offering a framework to explore the Hawking effect and the quantum properties of black holes \cite{wux2023, Haddadi:2023rjk}. Exotic Dilaton black holes, closely associated with string theory, black hole physics, and loop quantum gravity, have garnered significant attention. Investigating entanglement in the context of a Dilaton black hole is also widely regarded as a promising approach to enhancing our understanding of black holes, quantum gravity, and quantum information theory \cite{wucai2022}.\\

\subsubsection{Schwarzschild Black Hole}

The Schwarzschild black hole metric is given by \cite{brill1957, mm2010, shuminwu2023, Wu:2024jgy, Dolatkhah:2024dbm}
\begin{equation}
\label{Sch_BH}
	ds^2 = -c^2\left(1 - \frac{2GM}{c^2r}\right) dt^2 + \left(1 - \frac{2GM}{c^2r}\right)^{-1} dr^2 + r^2 d\Omega^2,
\end{equation}
where \( M \) is the mass of the black hole, \( r \) is the radial coordinate, and \( d\Omega^2 \) represents the line element of the unit sphere. For convenience, we set the gravitational constant \( G \), the Planck constant \( \hbar \), the Boltzmann constant \( k_B \), and the speed of light \( c \) to unity in this paper.

In a general spacetime background, the massless Dirac equation reads \cite{brill1957}
\begin{equation}
\label{Dirac_eqn}	
\left[\gamma^a e_{a}^{\mu} \left( \partial_{\mu} + \Gamma_{\mu} \right)\right] \Phi = 0,
\end{equation}
where \( \gamma^a \) are the Dirac matrices, \( \Gamma_{\mu} \) is the spin connection coefficient, and \( e_a^{\mu} \) is the inverse of the tetrad \( e_\mu^a \).

For the Schwarzschild spacetime, the Dirac equation (\ref{Dirac_eqn}) can be written explicitly as:
\begin{eqnarray}
	\label{Dirac_eqn_Sch}
	- \frac{\gamma_0}{\sqrt{1 - \frac{2M}{r}}} \frac{\partial \Phi}{\partial t} + \gamma_1 \sqrt{1 - \frac{2M}{r}}\left[\frac{\partial}{\partial r} + \frac{1}{r} + \frac{M}{2r\left(r-2M\right)}\right] \Phi + \frac{\gamma_2}{r}\left(\frac{\partial}{\partial \theta} + \frac{\cot \theta}{2}\right)\Phi + \frac{\gamma_3}{r \sin \theta}\frac{\partial \Phi}{\partial \phi} = 0,
\end{eqnarray}
where \( \gamma_i \) (\( i = 0, 1, 2, 3 \)) are the Dirac matrices \cite{Alsing:2006cj}. 

Solving the Dirac equation (\ref{Dirac_eqn_Sch}), we obtain the positive-frequency outgoing solutions for both the inside and outside regions of the event horizon \cite{Wang:2009zt, Xu:2014tza}:
\begin{eqnarray}
	\label{Dirac_out1}
	\Phi_{k,out}^{+} &\sim& \phi(r) e^{-i \omega u},\\
	\label{Dirac_out2}
	\Phi_{k,in}^{+} &\sim& \phi(r) e^{i \omega u},
\end{eqnarray}
where \( \phi(r) \) is the four-component Dirac spinor, \( u = t - r_* \) is the retarded time, and \( v = t + r_* \) is the advanced time, with the tortoise coordinate \( r_* = r + 2M \ln \left( \frac{r - 2M}{2M} \right) \). The frequency \( \omega \) is monochromatic, and \( k \) is the wave vector associated with the modes.

Thus, the Dirac field \( \Phi \) can be expanded as:
\begin{eqnarray}
\label{Dirac_out_Sch}
\Phi = \int dk \left[\hat{a}_k^{in} \Phi_{k,in}^{+} + \hat{b}_{-k}^{in \dagger} \Phi_{-k,in}^{-} + \hat{a}_k^{out} \Phi_{k,out}^{+} + \hat{b}_{-k}^{out \dagger} \Phi_{-k,out}^{-} \right],
\end{eqnarray}
where \( \hat{a}_k^{in} \) is the fermionic annihilation operator inside the event horizon, and \( \hat{b}_{-k}^{out \dagger} \) is the antifermionic creation operator outside the event horizon.

To describe the black hole using Kruskal coordinates, we introduce the generalized light-like coordinates \( U \) and \( V \) in Schwarzschild spacetime:
\[
u = -4M \ln\left[- \frac{U}{4M}\right], \quad v = 4M \ln\left[\frac{V}{4M}\right] \quad \text{for} \quad r > r_+,
\]
\[
u = -4M \ln\left[\frac{U}{4M}\right], \quad v = 4M \ln\left[\frac{V}{4M}\right] \quad \text{for} \quad r < r_+.
\]
Where, $r_+$ is the radius of event horizon.
Following the suggestion of Damour and Ruffini \cite{Damour:1976jd}, we can construct a complete basis for the positive-energy mode, or Kruskal mode, by analytically extending the solutions in equations (\ref{Dirac_out1}) and (\ref{Dirac_out2}). The Kruskal modes are given by:
\begin{eqnarray}
\label{kruskal_mode1}	
\Psi^{+}_{k,out} = e^{-2\pi M \omega} \Phi_{-k,in}^{-} + e^{2\pi M \omega} \Phi_{k,out}^{+},\\
 \label{kruskal_mode2}
\Psi^{+}_{k,in} = e^{-2\pi M \omega} \Phi_{-k,out}^{-} + e^{2\pi M \omega} \Phi_{k,in}^{+}.
\end{eqnarray}

Quantizing the Dirac field in terms of Kruskal modes, we have:
\begin{eqnarray}
	\label{Kruskal_mode_Sch}
	\Phi= \int dk\frac{1}{\sqrt{2 \cosh(4 \pi M \omega)}}\left[\hat{c}_{k}^{in} \Psi_{k,in}^{+} + \hat{d}_{-k}^{in \dagger} \Psi^{-}_{-k,in} + \hat{c}_{k}^{out} \Psi_{k,out}^{+} + \hat{d}_{-k}^{out \dagger} \Psi^{-}_{-k,out} \right],
\end{eqnarray}
where \( \hat{c}_k^{\sigma} \) and \( \hat{d}_k^{\sigma \dagger} \) with \( \sigma = (in, out) \) are the fermionic annihilation operators and antifermionic creation operators acting on the Kruskal vacuum.

From equations (\ref{Dirac_out_Sch}) and (\ref{Kruskal_mode_Sch}), it follows that the Dirac field is decomposed into both Schwarzschild and Kruskal modes. Using the Bogoliubov transformations, we can relate the Schwarzschild and Kruskal operators:
\begin{eqnarray}
	\hat{c}_{k}^{out}= \frac{1}{\sqrt{e^{- 8 \pi M \omega}+1}}\hat{a}_{k}^{out} - \frac{1}{\sqrt{e^{ 8 \pi M \omega}+1}} \hat{b}_{-k}^{in \dagger},\\
	\hat{c}_{k}^{out \dagger}= \frac{1}{\sqrt{e^{- 8 \pi M \omega}+1}}\hat{a}_{k}^{out \dagger} - \frac{1}{\sqrt{e^{ 8 \pi M \omega}+1}} \hat{b}_{-k}^{in}.
\end{eqnarray}

Using these transformations, we express the Kruskal vacuum and excited states in Schwarzschild spacetime as follows:
\begin{eqnarray}
\label{Kruskal_vac_Sch}
\vert 0 \rangle_K &=& \frac{1}{\sqrt{e^{\frac{-\omega}{T}} + 1}}\vert 0\rangle_{out} \vert 0\rangle_{in} + \frac{1}{\sqrt{e^{\frac{\omega}{T}} + 1}}\vert 1\rangle_{out} \vert 1\rangle_{in},\\
\label{exited_state krus}
\vert 1\rangle_{K} &=& \vert 1\rangle_{out} \vert 0\rangle_{in},
\end{eqnarray}
where \( |n\rangle_{out} \) and \( |n\rangle_{in} \) denote the fermionic modes outside and the antifermionic modes inside the event horizon, respectively, and \( T = \frac{1}{8 \pi M} \) is the Hawking temperature.

\subsubsection{Dilaton Black hole}
\noindent Using a similar method to that applied to static and asymptotically flat black holes, such as the Schwarzschild spacetime black hole, we derive the vacuum structure for Dirac particles in the background of a Garfinkle–Horowitz–Strominger (GHS) Dilaton black hole. The metric corresponding to a static, spherically symmetric, and charged Dilaton black hole can be expressed as \cite{shahbazi2010,Malik2024,smwu2022,Damour:1976jd,Garfinkle:1990qj,Garcia:1995qz}
\begin{eqnarray}
\label{Dilaton_BH}
    ds^2 = -\left(\frac{r-2M}{r-2D}\right) dt^2 + \left(\frac{r-2M}{r-2D}\right)^{-1} dr^2 + r\left(r-2D\right) d \Omega^2
\end{eqnarray}
where $M$ is the black hole mass, $D$ is the Dilaton of the black hole and $d \Omega^2$ is the line element in the unit sphere. The relationship between the mass $M$, the charge $Q$, and the Dilaton $D $ is given by $D = \frac{Q^2}{2M}$. Additionally, the Dilaton $D$ and the mass $M$ of the black hole must satisfy $D < M$.
Now, the massless Dirac equation (\ref{Dirac_eqn}) in the GHS Dilaton black-hole spacetime becomes
\begin{eqnarray}
\label{Dirac_eqn_Dila}
    - \frac{\gamma_0}{\sqrt{f}} \frac{\partial \Phi}{\partial t} + \gamma_1 \sqrt{f}\left[\frac{\partial}{\partial r} + \frac{r-D}{r(r-2D)} + \frac{1}{4f} \frac{d f}{d r}\right] \Phi + \frac{\gamma_2}{\sqrt{r (r-2D)}}\left(\frac{\partial}{\partial \theta} + \frac{\cot \theta}{2}\right)\Phi + \frac{\gamma_3}{\sqrt{r (r-2D)} \sin \theta}\frac{\partial \Phi}{\partial \phi} = 0,
\end{eqnarray}
where $f=\frac{r-2M}{r-2D}$. Then, by solving Dirac equation in the GHS Dilaton spacetime, we can obtain the positive frequency outgoing solutions outside and inside regions of the event horizon
\begin{eqnarray}
\label{Dirac_out11}
    \Phi_{k,out}^{+} &\sim& \mathcal{R} e^{-i \omega u},\\
    \label{Dirac_out22}
    \Phi_{k,in}^{+} &\sim& \mathcal{R} e^{i \omega u},
\end{eqnarray}
where $\mathcal{R}$ is the four-component Dirac spinor, retarded time $u = t - r_{\ast}$ and advanced time $v = t + r_{\ast}$ with the tortoise coordinate $r_{\ast}=r+2(M-D)\ln{\frac{r-2M}{2M-2D}}$. Now, we can expand Dirac field through equations (\ref{Dirac_out11}) and (\ref{Dirac_out22}) as 
\begin{eqnarray}
\label{Dirac_Field_Dila}
     \Phi = \int dk\left[\hat{a}_{k}^{in}\phi_{k,in}^{+} + \hat{b}_{-k}^{in \dagger} \Phi_{-k,in}^{-} + \hat{a}_{k}^{out}\phi_{k,out}^{+} + \hat{b}_{-k}^{out \dagger} \Phi_{-k,out}^{-} \right],
\end{eqnarray}
where $\hat{a}_k^{in}$ is the fermionic annihilation operator inside the event horizon, while $\hat{b}^{out \dagger}_{-k}$ is the antifermionic creation operator outside the event horizon of the black hole.\\
\noindent The generalized Kruskal coordinates $U$ and $V$ for the GHS Dilaton spacetime can be arranged as 
\begin{eqnarray}
    u&=&-4(M - D) \ln\left[-\frac{U}{4(M-D)}\right],~~ 
    v= 4(M - D) \ln\left[\frac{V}{4(M-D)}\right]~~for~ r > r_{+}~; \nonumber\\
    u &=& -4(M - D) \ln\left[\frac{U}{4(M-D)}\right],~~ 
    v = 4(M - D) \ln\left[\frac{V}{4(M-D)}\right]~~for~ r < r_{+}~. \nonumber
\end{eqnarray}
Using the relation between Kruskal and black hole coordinates, we can derive a complete basis of positive energy modes which are analytic by making ana analytic continuations for equations (\ref{Dirac_out11}) and (\ref{Dirac_out22}) according to the suggestions of Damoar and Ruffini \cite{Damour:1976jd},
\begin{eqnarray}
    \label{kruskal_mode11}
    \Psi^{+}_{k,out} = e^{-2 (M-D) \pi \omega} \Phi_{-k,in}^{-} + e^{2 \pi (M-D) \omega} \Phi_{k,out}^{+} \\
    \label{kruskal_mode22}
    \Psi^{+}_{k,in} = e^{-2 \pi (M-D) \omega} \Phi_{-k,out}^{-} + e^{2 \pi (M-D) \omega} \Phi_{k,in}^{+} .
\end{eqnarray}\\
Now, we can quantize the Dirac field in Kruskal modes
\begin{eqnarray}
    \label{Kruskal_mode_Dia}
    \Phi= \int dk\frac{1}{\sqrt{2 \cosh(4 \pi (M-D) \omega)}}\left[\hat{c}_{k}^{in} \Psi_{k,in}^{+} + \hat{d}_{-k}^{in \dagger} \Psi^{-}_{-k,in} + \hat{c}_{k}^{out} \Psi_{k,out}^{+} + \hat{d}_{-k}^{out \dagger} \Psi^{-}_{-k,out} \right],
\end{eqnarray}
where $\hat{c}_{k}^{\sigma}$ and $\hat{d}_{k}^{\sigma \dagger}$ with $\sigma = (in,~out)$ are the fermionic annihilation operators and antifermionic  creation operators that act on Kruskal vacuum. From equations (\ref{Dirac_Field_Dila}) and (\ref{Kruskal_mode_Dia}), it is clear that the Dirac field is decomposed into the GHS Dilaton and Kruskal modes, respectively. Now, one can easily evaluate the Bogoliubov transformations between annihilation and creation operators in GHS Dilaton  and Kruskal coordinates. Applying the Bogoliubov transformations, GHS Dilaton and Kruskal operators take the forms
\begin{eqnarray}
\hat{c}_{k}^{out}= \frac{1}{\sqrt{e^{- 8 \pi (M-D) \omega}+1}}\hat{a}_{k}^{out} - \frac{1}{\sqrt{e^{ 8 \pi (M-D) \omega}+1}} \hat{b}_{-k}^{in \dagger},\\
\hat{c}_{k}^{out \dagger}= \frac{1}{\sqrt{e^{- 8 \pi (M-D) \omega}+1}}\hat{a}_{k}^{out \dagger} - \frac{1}{\sqrt{e^{ 8 \pi (M-D) \omega}+1}} \hat{b}_{-k}^{in}.
\end{eqnarray}
The spacetime of the GHS Dilaton black hole can be divided into physically inaccessible and accessible regions. The ground state mode in GHS Dilaton black hole coordinates corresponds to a two-mode squeezed state in Kruskal coordinates. After suitably normalizing the state vector, the expressions of the Kruskal vacuum and excited states can be expressed as
\begin{eqnarray}
    \label{Kruskal_vac_Dia}
    \vert 0 \rangle_K &=& \frac{1}{\sqrt{e^{ - 8 \pi (M-D) \omega}+1}}\vert 0\rangle_{out} \vert 0\rangle_{in} + \frac{1}{\sqrt{e^{ 8 \pi (M-D) \omega}+1}}\vert 1\rangle_{out} \vert 1\rangle_{in},\\
    \label{exited_state}
    \vert 1\rangle_{K} &=& \vert 1\rangle_{out} \vert 0\rangle_{in},
\end{eqnarray}
where $\vert n \rangle_{out}$ and $\vert n \rangle_{in}$ represent the fermionic modes outside the event horizon and the antifermionic modes inside the event horizon, respectively.

\subsection{Quantification of genuine tripartite entanglement of tripartite states exposed to Schwarzschild and Dilaton Black hole}
\subsubsection*{Tangle: Overview of the measure:}
\noindent There is a measure by which we can quantify genuine tripartite entanglement among the parties holding qubits of a state. This measure is known as \textit{tangle} \cite{coffman2000}. The tangle of a tripartite state is quantified as
 \begin{eqnarray}
     \label{tangle}
     \tau_{ABC} = C^{2}_{A(BC)} - C^{2}_{AB} - C^{2}_{AC}.
 \end{eqnarray} 
 Here $C$ is a well-known measure to quantify entanglement in bipartite system. $C_{A(BC)}$ is the concurrence between party $A$ and joint state $BC$, likewise $C_{AB}$ (and $C_{AC}$) are the concurrences quantifying entanglement of between parties $A$ and $B$ ($A$ and $C$). The concurrence of a bipartite quantum state $\rho$ \cite{wootters1998} is defined as
\begin{eqnarray}
\label{concurrence}
C(\rho) = \max \lbrace 0, \sqrt{\lambda_{1}}-\sqrt{\lambda_{2}}-\sqrt{\lambda_{3}}-\sqrt{\lambda_{4}}\rbrace,  
\end{eqnarray}
where $\lambda_{1}\ge\lambda_{2}\ge\lambda_{3}\ge\lambda_{4}$ are the eigenvalues of the matrix $\rho \tilde{\rho}$.  The spin-flipped density matrix $\tilde{\rho}$ is
\begin{eqnarray}
\label{spin-flipped}
\tilde{\rho} = (\sigma_{y}\otimes \sigma_{y})\rho^{*}(\sigma_{y}\otimes \sigma_{y}),
\end{eqnarray}
where $\sigma_{y}$ is the Pauli spin matrix in the $y$-basis and $\tilde{\rho}$ is in the same basis as $\rho$, and  $\rho^{*}$ is the complex conjugate of the density matrix $\rho$. 
\subsubsection{Tripartite states of two in-equivalent classes:}
\noindent We have already mentioned in the sec.\ref{sec:introduction} that there are two in-equivalent class of states which are $GHZ$ and $W$ states. The states are defined below.
\begin{eqnarray}
\label{ghzpure}
    \vert GHZ\rangle_{ABC} &=& \frac{1}{\sqrt{2}}\Big(\vert 000\rangle_{ABC} + \vert 111\rangle_{ABC}\Big),
    \end{eqnarray}
and
\begin{eqnarray}
\label{wpure}
 \vert W\rangle_{ABC} &=& \frac{1}{\sqrt{3}}\Big(\vert 100\rangle_{ABC} + \vert 010\rangle_{ABC} + \vert 001\rangle_{ABC}\Big).
\end{eqnarray}
The tangle ($\tau$) of these states respectively are $1$ and $0$ \cite{coffman2000}. However, although $GHZ$ states are useful as quantum teleportation channels, $W$ states are not \cite{pati2006}. The $W$ states that we defined above in eq.(\ref{wpure}) can be mentioned as prototypical $W$ states. Rather, Agrawal and Pati \cite{pati2006} worked with non-prototypical $W$ states defined as
\begin{eqnarray}
\label{wpure1}
\vert W_{1}\rangle_{ABC} = \frac{1}{2}\Big[\vert 100\rangle_{ABC} + \vert 010\rangle_{ABC} + \sqrt{2}\vert 001\rangle_{ABC}\Big],
\end{eqnarray}
and have shown that they could perform perfect teleportation with the state defined in eq.(\ref{wpure1}) while the tangle ($\tau$) of this non-prototype $W$ state is also $0$. Also it is a known fact $GHZ$ state loses its entanglement in bipartite level i.e. $C(\rho_{GHZ}^{AB}),\:C(\rho_{GHZ}^{BC}),\:C(\rho_{GHZ}^{AC})$ is $0$, after removal of any party $A$, $B$ or $C$. This means that such bipartite mixed states derived from $GHZ$ state is not suitable as quantum teleportation channel. Again, when the non-prototypical $W$ state $W_{1}$ of eq.(\ref{wpure1}) is taken into consideration, and subsequently one of the parties $A$, $B$ or $C$ are removed, then using eq.(\ref{concurrence}) we see that $C(\rho_{W1}^{AB})=0.50$ while $C(\rho_{W1}^{AC})=C(\rho_{W1}^{BC})=0.207$, where $\rho_{W1}$ is the density matrix corresponding to the state $\vert W_{1}\rangle_{ABC}$. Although the tripartite $W$ state of eq.(\ref{wpure}) is not useful as quantum channel for perfect teleportation, the state has high bipartite entanglement than that of $W_{1}$ state and the concurrence of such bipartite state $\rho_{W}$, corresponding to the state $W$, is found as $C(\rho_{W}^{AB})=C(\rho_{W}^{AC})=C(\rho_{W}^{BC})=0.67$ \cite{roy2023}.\\\\

\subsubsection{Tripartite states subjected to black holes:}
\noindent Below we represent a diagram of the scenario that is our focus of the current study.
\begin{figure}[h]
\label{schematic}
\includegraphics[width=8.7cm]{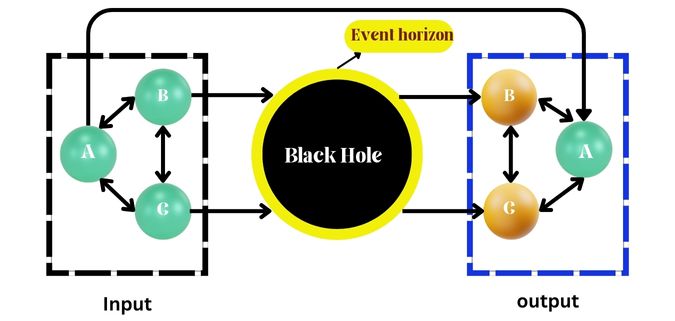}
\caption{The schematic diagram shows that the $3$ qubit states being exposed near the event horizon of the  black hole.}
\label{schematic}
\end{figure}
In the figure \ref{schematic}, a schematic diagram shows tripartite states held by Alice (A, flat region), Bob (B), and Cliff (C) near Schwarzschild/Dilaton event horizon($r_h$). ``Near-horizon'' for B/C means fixed $r \gtrsim r_h$ ($r_h=2M$ Schwarzschild, $r_h=2M-2D$ Dilaton), where local Dirac modes mix via Hawking effect (eqs.(\ref{Dirac_out_Sch})--(\ref{Kruskal_mode_Sch}), (\ref{Dirac_Field_Dila})--(\ref{Kruskal_mode_Dia})). Particles propagate on outgoing null geodesics $e^{-i\omega(t-r_*)}$, but entanglement computed at snapshot before full infall ($\tau_\text{cross} \sim r_h \ln[(r-r_h)/r_h]$ proper time). There may or may not be genuine tripartite entanglement among parties as in the cases of $GHZ$ and $W$ states. The rectangular box with dashed border represent the physical system of the tripartite state. The yellow circular region surrounding the black hole depicts the event horizon. The left hand side region of the black hole shows the physical system of the tripartite state before the parties Bob and Cliff are exposed to event horizon whereas the right hand side of the black hole represent the physical system of the same tripartite state after Bob and Cliff are exposed to event horizon of the black hole. Throughout the transition Alice remains in the flat region.
\subsubsection{GHZ state in Black Hole}
\noindent We consider the $GHZ$ state as is defined in eq.(\ref{ghzpure}) and we assume that initially the qubits are distributed among parties Alice ($A$), Bob ($B$) and Cliff ($C$). Also we consider that Alice stays in the flat (stationary) region while Bob and Cliff are near the event horizon of (i)Schwarzschild black hole and (ii) Dilaton Black Hole.\\\\
To obtain the wave function in the curved spacetime background, we must map the states of the observers near the event horizon (Bob and Cliff) from the Minkowski basis to the Kruskal basis. This mapping is derived from the Kruskal mode expansions given in eqs. (\ref{Kruskal_vac_Sch}) and (\ref{exited_state krus}), where we express the Kruskal vacuum and excited states as superpositions of the Schwarzschild outer and inner horizon states. The construction of the full wavefunction follows a systematic substitution procedure: (i) The flat Minkowski vacuum state $|0\rangle$ for a party near the horizon is replaced by the Kruskal vacuum: $|0\rangle \rightarrow |0\rangle_K = \mu|0\rangle_{out}|0\rangle_{in} + \nu|1\rangle_{out}|1\rangle_{in}$, and (ii) The flat Minkowski excited state $|1\rangle$ for a party near the horizon is replaced by the Kruskal single-particle state: $|1\rangle \rightarrow |1\rangle_K = |1\rangle_{out}|0\rangle_{in}$. Applying these substitutions term by term to the original $GHZ$ state of eq. (\ref{ghzpure}) yields:
\begin{eqnarray}
|GHZ\rangle_{ABC} &=& \frac{1}{\sqrt{2}}\big(|0\rangle_A|0\rangle_B|0\rangle_C + |1\rangle_A|1\rangle_B|1\rangle_C\big) \nonumber \\
&\xrightarrow{\text{substitution}}& \frac{1}{\sqrt{2}}\Big[ |0\rangle_A \big(\mu|0\rangle_{B_{out}}|0\rangle_{B_{in}} + \nu|1\rangle_{B_{out}}|1\rangle_{B_{in}}\big) \otimes \big(\mu|0\rangle_{C_{out}}|0\rangle_{C_{in}} + \nu|1\rangle_{C_{out}}|1\rangle_{C_{in}}\big) \nonumber \\
&&\quad + |1\rangle_A \big(|1\rangle_{B_{out}}|0\rangle_{B_{in}}\big) \otimes \big(|1\rangle_{C_{out}}|0\rangle_{C_{in}}\big) \Big]. \nonumber
\end{eqnarray}
Expanding this expression and rearranging terms in the order $A$, $B_{out}$, $\bar{B}\equiv B_{in}$, $C_{out}$, $\bar{C}\equiv C_{in}$ yields precisely wave function of $GHZ$ state as 

\begin{eqnarray}
\label{ghz1}
\vert GHZ\rangle_{wf} &=& \frac{1}{\sqrt{2}}\Big(\mu^2 \vert 00000\rangle_{AB\bar{B}C\bar{C}} + \mu\nu \vert  00011\rangle_{AB\bar{B}C\bar{C}}\nonumber\\ + && \mu\nu \vert  01100\rangle_{AB\bar{B}C\bar{C}} +  \nu^2 \vert 01111\rangle_{AB\bar{B}C\bar{C}} + \vert 11010\rangle_{AB\bar{B}C\bar{C}}\Big).
\end{eqnarray}
Here $\bar{B}$ and $\bar{C}$ are inaccessible modes of Bob and Cliff and we should trace out these parts. Thus tracing out $\bar{B}$ and $\bar{C}$ from eq.(\ref{ghz1}), we get density matrix as
\begin{eqnarray}
\label{ghz2}
\rho_{GHZ}^{wfabc}=\frac{1}{2}\begin{pmatrix}
\mu^4  & 0 & 0 & 0& 0 & 0 & 0 & \mu^2\\
0  & \mu^2\nu^2 & 0 & 0& 0 & 0 & 0 & 0\\
0  & 0 & \mu^2\nu^2 & 0& 0 & 0 & 0 & 0\\
0  & 0 & 0 & 0& 0 & 0 & 0 & 0\\
0  & 0 & 0 & 0& \nu^4 & 0 & 0 & 0\\
0  & 0 & 0 & 0& 0 & 0 & 0 & 0\\
0  & 0 & 0 & 0& 0 & 0 & 0 & 0\\
\mu^2  & 0 & 0 & 0& 0 & 0 & 0 & 1\\
\end{pmatrix}.
\end{eqnarray}
Now using eq.(\ref{tangle}), we find
\begin{eqnarray}
\label{tangleghz1}
\tau(\vert GHZ\rangle_{ABC}) = \mu^2 + \nu^2.
\end{eqnarray}
In case, (i) the parties Alice, Bob and Cliff share $GHZ$ state in Schwarzschild black hole, we have 
$\mu = \frac{1}{\sqrt{e^{\frac{-\omega}{T}} + 1}}$ and $\nu = \frac{1}{\sqrt{e^{\frac{\omega}{T}} + 1}}$ while (ii) the parties Alice, Bob and Cliff share $GHZ$ state in Dilaton black hole, we have $\mu = \frac{1}{\sqrt{e^{ - 8 \pi (M-D) \omega}+1}}$ and $\nu = \frac{1}{\sqrt{e^{ 8 \pi (M-D) \omega}+1}}$.
\subsubsection*{Results and discussion:}
\noindent We plot below the tangle of $GHZ$ state which is found to be $\mu^2 + \nu^2$ shown in eq.(\ref{tangleghz1}).\begin{figure}[h]
\includegraphics[width=8.7cm]{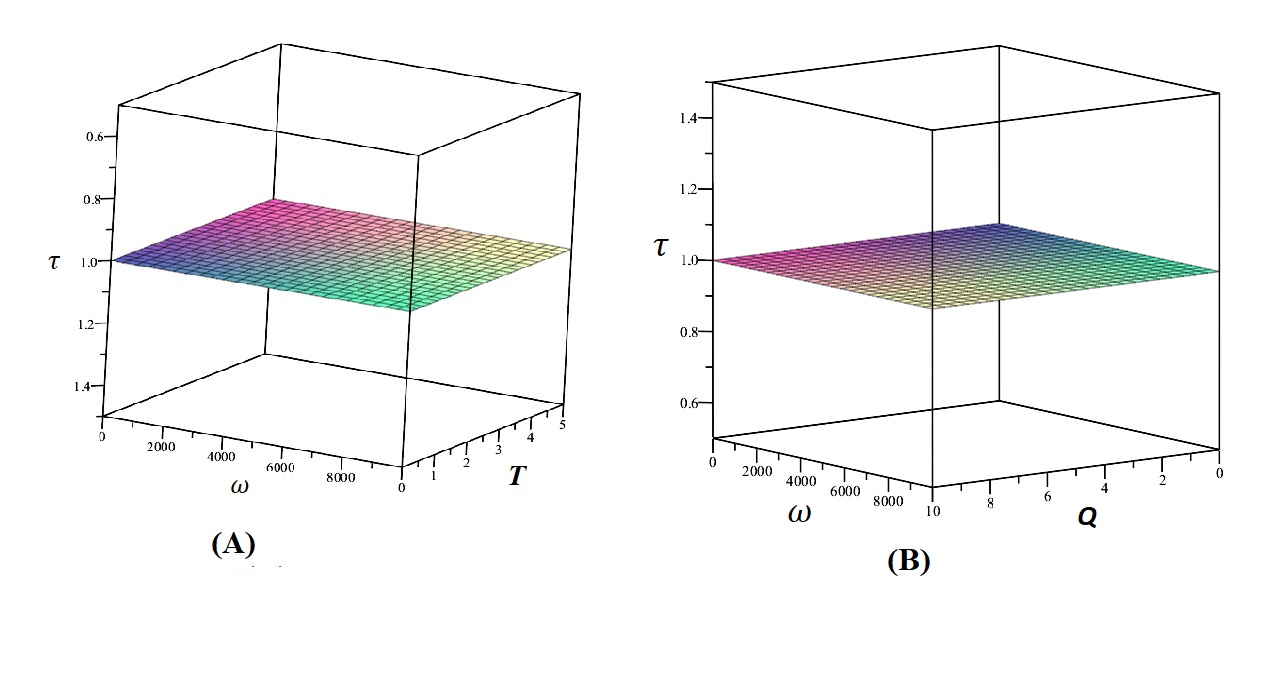}
\label{fig2}
\caption{We plot the tangle of the $GHZ$ state against monochromatic frequency ($\omega$) and Hawking temperature ($T$). Fig.(A) shows the variation of tangle $\tau(\rho_{GHZ}^{wfabc})$ when Hawking temperature ($T$) is varied from $0$ to $5$ and monochromatic frequency ($\omega$) is varied from $0$ to $10000$ in the background of Schwarzschild black hole. Fig.(B) shows the variation of same $\tau(\rho_{GHZ}^{wfabc})$ when charge ($Q$) is varied from $0$ to $20$ and monochromatic frequency ($\omega$) is varied from $0$ to $50$ in the \textcolor{blue}{background} of Dilaton black hole.}
\label{fig2}
\end{figure} 
In figure \ref{fig2}, we plot the tangle ($\tau$) of $GHZ$ state as a function of the Hawking temperature ($T$) and monochromatic frequency ($\omega$), in the background of Schwarzschild black hole. We observe that the entanglement of the $GHZ$ state reaches the value of $1$ and remains at this value $1$ when the monochromatic frequency becomes sufficiently large. Here we use tangle as the measure of showing genuine tripartite entanglement. It is established that the tangle of $GHZ$ state equals $1$ in the flat region. 
In fig \ref{fig2}(B), we analyze the variation of the tangle of the $GHZ$ state as a function of the charge ($Q$) and monochromatic frequency ($\omega$) in the context of a Dilaton black hole space time.Our findings indicate that the tangle of the $GHZ$ state consistently remains at a value of $1$. \\

As we know that the $GHZ$ states have no bipartite entanglement left in them, it is quite expected that the parties holding the qubits of $GHZ$ state, when exposed to the regions as shown in the schematic diagram fig.\ref{schematic}, will have no bipartite entanglement left in the state either. 
Using eq.(\ref{concurrence}), it can easily be seen that $C(\rho_{GHZ}^{wfbc})=C(\rho_{GHZ}^{wfac})=C(\rho_{GHZ}^{wfab})=0$.
\subsection{$W$ states in Black Hole}
\noindent In this section we consider prototype and non-prototype $W$ states in the background of Schwarzschild and Dilaton black holes. 
\subsubsection{Protoype $W$ state in Black hole:}
We consider the $W$ state as is defined in eq.(\ref{wpure}) and we assume that initially the qubits are distributed among parties Alice ($A$), Bob ($B$) and Cliff ($C$). Alice stays in the flat (stationary) region while Bob and Cliff are near the event horizon of these two black holes. 
Following the same substitution procedure detailed in Section II.2.3 for the $GHZ$ state, we apply the Kruskal mode expansion to the prototype $W$ state of eq.(\ref{wpure}). Replacing $|0\rangle_B$, $|1\rangle_B$, $|0\rangle_C$, and $|1\rangle_C$ with their Kruskal counterparts as given in eqs.(\ref{Kruskal_vac_Sch}) and (\ref{exited_state krus}), and expanding term by term, we obtain the wave function of $W$ state as:

\begin{eqnarray}
\label{protow}
\vert W\rangle_{wf} &=& \frac{1}{\sqrt{3}}\Big[\mu\vert 00010\rangle_{AB\bar{B}C\bar{C}} + \nu\vert 01110\rangle_{AB\bar{B}C\bar{C}} + \mu\vert 01000\rangle_{AB\bar{B}C\bar{C}} + \nu\vert 01011\rangle_{AB\bar{B}C\bar{C}}+\nonumber\\&& + \mu^2\vert 10000\rangle_{AB\bar{B}C\bar{C}} + \mu\nu\vert 10011\rangle_{AB\bar{B}C\bar{C}} + \mu\nu\vert 11100\rangle_{AB\bar{B}C\bar{C}} + \nu^2\vert 11111\rangle_{AB\bar{B}C\bar{C}}\Big.]
\end{eqnarray}
Tracing out the inaccessible regions of $B$ and $C$ i.e. $\bar{B}$ and $\bar{C}$ we get
\begin{eqnarray}
\label{denw}
\rho_{W}^{wfabc}=\frac{1}{3}\begin{pmatrix}
0  & 0 & 0 & 0& 0 & 0 & 0 & 0\\
0  & \mu^2 & \mu^2 & \mu^3 & 0 & 0 & 0 & 0\\
0  & \mu^2 & \mu^2 & \mu^3& 0 & 0 & 0 & 0\\
0  & \mu^3 & \mu^3 & \mu^4 & 0 & 0 & 0 & 0\\
0  & 0 & 0 & 0& 2\nu^2 & \mu\nu^2 & \mu\nu^2 & 0\\
0  & 0 & 0 & 0& \mu\nu^2 & \mu^2\nu^2 & 0 & 0\\
0  & 0 & 0 & 0& \mu\nu^2 & 0 & \mu^2\nu^2 & 0\\
0  & 0 & 0 & 0& 0 & 0 & 0 & \nu^4\\
\end{pmatrix}.
\end{eqnarray}
In this work we are least interested in studying the genuine tripartite entanglement of $W$ state when state is exposed to Black hole regions, this is because the state has no genuine tripartite entanglement as $\tau(W) = 0$. But the state retains the bipartite entanglement and hence we shall remove parties $A$, $B$ and $C$ individually from the exposed $W$ state and will study the bipartite scenario. Since Alice ($A$) is in flat region, Alice is not disturbed while removing  the parties $B$ and $C$ (who are in contact with event horizon) are removed one at a time. Thus, we get following bipartite mixed states.
\begin{eqnarray}
\label{rhoabipartite2}
\rho_{W}^{wfac}&=&\rho_{W}^{wfab} =\frac{1}{3}\begin{pmatrix}
\mu^2  & 0 & 0 & 0\\
0  & \mu^2+\nu^2+\nu^4 & \mu^3+\mu\nu^2 & 0\\
0  & \mu^3+\mu\nu^2 & \mu^4+\mu^2\nu^2 & 0\\
0  & 0 & 0 &  \mu^2\nu^2+\nu^4 \\
\end{pmatrix}.
\end{eqnarray} 
Using eq.(\ref{concurrence}), we find that concurrence of the bipartite mixed state when parties $B$ and $C$ are traced out from state (\ref{denw}) as
\begin{eqnarray}
\label{concwfac1}
C(\rho_{W}^{wfac}) &=& C(\rho_{W}^{wfab})=\sqrt{f_1}-\sqrt{f_2}-\sqrt{f_3}-\sqrt{f_4}.\nonumber\\
f_1&=&\frac{1}{9}\Big[\nu^4 + 2(\mu^2+\nu^2)+2\sqrt{(\mu^2+\nu^2)(\nu^4+\mu^2+\nu^2)}\Big](\mu^2+\nu^2)\mu^2,\nonumber\\
f_2&=&\frac{1}{9}\Big[\nu^4 + 2(\mu^2+\nu^2)-2\sqrt{(\mu^2+\nu^2)(\nu^4+\mu^2+\nu^2)}\Big](\mu^2+\nu^2)\mu^2,\nonumber\\
f_3&=& f_4 =\frac{1}{9}\mu^2\nu^2\Big(\mu^2+\nu^2\Big),
\end{eqnarray}\\\\
The concurrence of the state $\rho_{W}^{wfac}$ (or \textcolor{blue}{$\rho_{W}^{wfab}$}) are plotted against the monochromatic frequency ($\omega$) and Hawking temperature ($T$) in the following figure.
\begin{figure}[h]
\label{conc2plot}
\includegraphics[width=8.7cm]{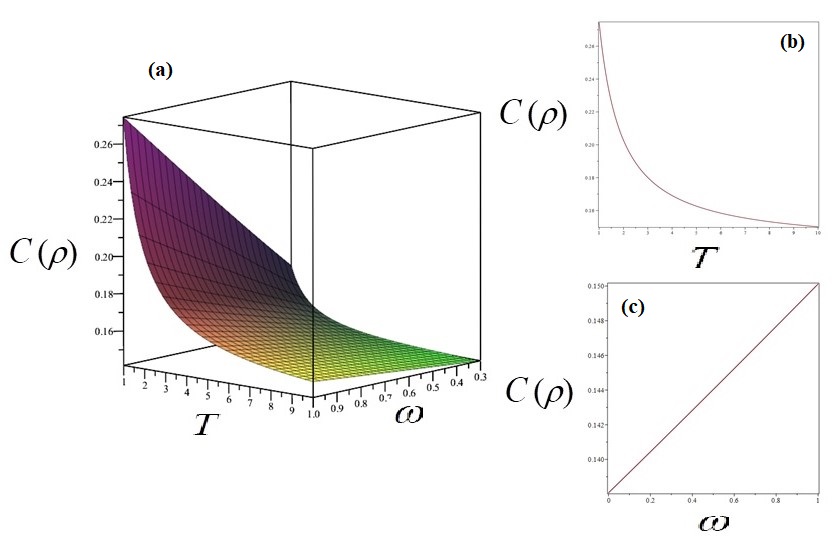}
\caption{Variation of concurrence of bipartite mixed states $\rho_{W}^{wfac}$ (or $\rho_{W}^{wfab}$) derived from tripartite prototype $W$ states in the background of Schwarzschild black hole. Fig.(a) is the $3D$ depiction of the variation of $C(\rho_{W}^{wfac})$ against monochromatic frequency ($\omega$) and Hawking temperature ($T$), while $\omega$ is varied from $0$ to $1$, $T$ is varied from $1$ to $10$. Fig.(b) is the $2D$ plot of concurrence against $T$ and fig.(c) is the $2D$ plot concurrence against $\omega$.}
\label{conc2plot}
\end{figure}\\\\
It is clear from fig.\ref{conc2plot} that when monochromatic frequency is $\omega = 1$ (i.e. at the upper limit as per our choice) and Hawking temperature $T=1$ (i.e. at the lower limit as per our choice) the concurrence $C(\rho_{W}^{wfac})$ (or of $C(\rho_{W}^{wfab})$) is greater than $0.26$. With increase in Hawking temperature concurrence is decreasing and on the other hand with increase in monochromatic frequency there is an increase in concurrence of this bipartite mixed state (see figs.\ref{conc2plot}(b) and \ref{conc2plot}(c)). Bipartite entanglement of the $W$ state with respect to measure of concurrence was found to be $0.67$, but here this low value of concurrence in the bipartite counterpart of $W$ state is quite justified as the state was under the impact of Black hole's event horizon. That means it loses some entanglement in the context of concurrence due to the influence of event horizon on qubits held by Bob and Cliff.\\\\
Again we see how concurrence of these bipartite mixed states fluctuates in the background of Dilaton parameters in the following figure.
\begin{figure}[h]
\label{conc2plotb}
\includegraphics[width=8.7cm]{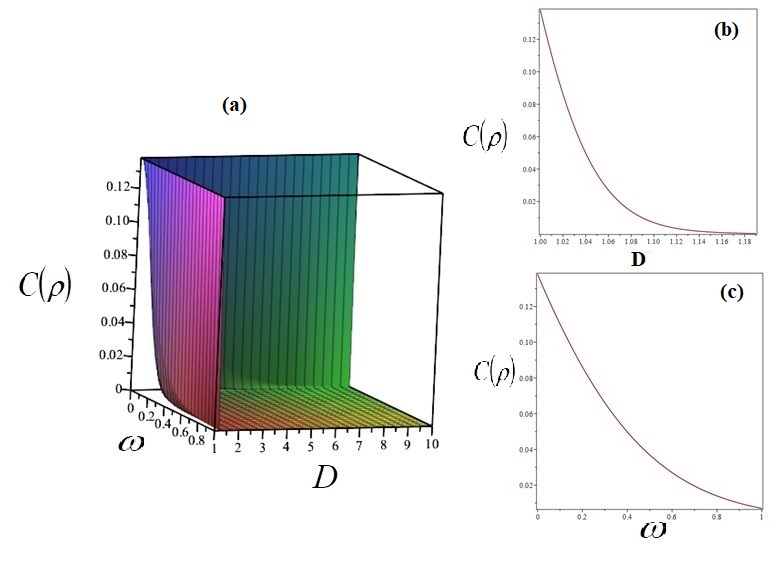}
\caption{Variation of concurrence of bipartite mixed states $\rho_{W}^{wfac}$ (or $\rho_{W}^{wfab}$) derived from tripartite prototype $W$ states in the background of Dilaton black hole. Fig.(a) is the $3D$ depiction of the variation of $C(\rho_{W}^{wfac})$ against monochromatic frequency ($\omega$) and Dilaton parameter ($D$), while $\omega$ is varied from $0$ to $1$, $D$ is varied from $1$ to $10$. Fig.(b) is the $2D$ plot of concurrence against $D$ and fig.(c) is the $2D$ plot concurrence against $\omega$.}
\label{conc2plotb}
\end{figure}
From fig.\ref{conc2plotb}, it is clear that, when the monochromatic frequency $\omega = 0$ and $D=1$, then $C(\rho_{W}^{wfac})>0.12$, while for $\omega = 1$ as well as for $D>1.16$, $C(\rho_{W}^{wfac})$ vanishes. We also observe that both with increase in the value of Dilaton parameter ($D$) and monochromatic frequency ($\omega$), concurrence vanishes quickly. In comparison to the situation observed in fig.\ref{conc2plot}, the value of concurrence here is too low, so that we can conclude that event horizon of Dilaton black hole has much more deteriorating impact on the concurrence of the bipartite mixed state $\rho_{W}^{wfac}$ (or $\rho_{W}^{wfab}$). \\\\
We shall now investigate whether the mixed states obtained above can be used as teleportation channels and this can only happen when the teleportation fidelity can exceed the classical teleportation fidelity $\frac{2}{3}$, which is the maximum fidelity achievable by means of local operations and classical communications \cite{pop1994,mas1995,gis1996}.\\\\ For a bipartite state $\rho$, when used as quantum teleportation channel, it's optimal teleportation fidelity ($f^T$) is given by
\begin{eqnarray}
\label{telepfid}
f^{T}(\rho) = \frac{1}{2}\Big\lbrace 1+ \frac{N(\rho)}{3} \Big\rbrace,
\end{eqnarray}
where $N(\rho) = \sum_{i = 1}^{3}\sqrt{u_{i}}$. Here $u_{i}$'s are the eigenvalues of the matrix $T_{1}^{\dagger}T_{1}$. The elements of $T_{1}$ are denoted by $t_{nm}$ and these elements are calculated as $t_{nm} = Tr(\rho\sigma_{n}\otimes\sigma_{m})$ where $\sigma_{j}$'s denote the Pauli spin operators. In terms of teleportation fidelity, a general result holds that any mixed spin-$\frac{1}{2}$ state $\rho$ is useful for standard teleportation if and only if $N(\rho)>1$ \cite{horo1996}.\\\\
\textit{Teleportation fidelity of the state $\rho_{W}^{wfac}$ ($\rho_{W}^{wfab}$):} We know that when we consider the $W$ state and remove any one party from it, the resultant bipartite state is mixed. This mixed state can be successfully used as quantum teleportation channel, whereas, the teleportation fidelity of the state is $\frac{7}{9}$ exceeding the classical teleportation fidelity $\frac{2}{3}$\cite{roy2010,roy2023}. Now considering the bipartite mixed state (\ref{rhoabipartite2}), we calculate the teleportation fidelity using eq.(\ref{telepfid}) and see that
\begin{eqnarray}
\label{telepfidrhoabipartite2}
f^{T}(\rho_{W}^{wfac}) &=& f^{T}(\rho_{W}^{wfab})= \frac{1}{2}+\frac{1}{18}\Big(\mu^4 +\nu^2\Big) + \frac{2}{9}\Big(\mu^3+\mu\nu^2\Big).
\end{eqnarray}\\\\
Below we plot the teleportation fidelity of the state $\rho_{W}^{wfac}$ (or \textcolor{blue}{$\rho_{W}^{wfab}$})
against the parameters $\omega$ and $T$.
\begin{figure}[h]
\label{telfidw}
\includegraphics[width=8.7cm]{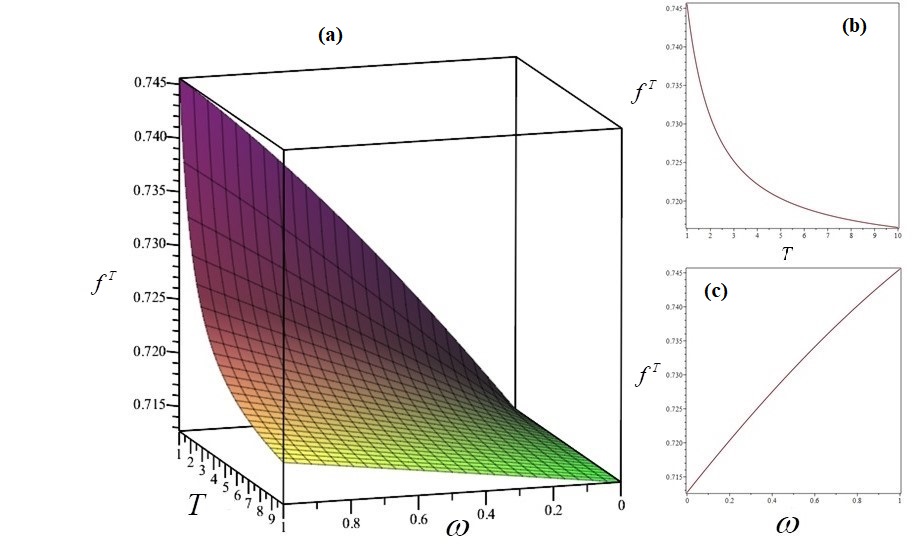}
\caption{Variation of teleportation fidelity of bipartite mixed states $\rho_{W}^{wfac}$ (or $\rho_{W}^{wfab}$) derived from tripartite prototype $W$ states in the background of Schwarzschild black hole. $T$ is varied from $1$ to $10$ and $\omega$ is varied from $0$ to $1$. Fig.(b) shows variation of teleportation fidelity against $T$ and fig. (c) shows variation of teleportation fidelity against $\omega$.}
\label{telfidw}
\end{figure}
\section{Results and discussion:}
\noindent From fig.\ref{telfidw} we see that,  teleportation fidelity of the states $\rho_{W}^{wfac}$ (and $\rho_{W}^{wfab}$) varies with Hawking temperature ($T$) as well as monochromatic frequency $\omega$. In the left hand side of the figure \ref{telfidw}(a), we have shown the variation of $f^{T}(\rho_{W}^{wfac})$ (or $f^{T}(\rho_{W}^{wfab})$) with respect to $\omega$ and $T$ while on the right side i.e. figs.\ref{telfidw}(b) and \ref{telfidw}(c), we show two dimensional overview. It is observed that when the value of the monochromatic frequency ($\omega$) is fixed at $1$ and subsequently Hawking temperature ($T$) is varies from $1$ to $10$, $f^{T}$ decreases from $0.745$ to $0.715$ (fig.\ref{telfidw}(b)) while the matter is just the opposite when Hawking temperature is fixed at $1$ and $\omega$ is varied from $0$ to $1$ (i.e. the teleportation fidelity then increases from $0.715$ to $0.745$)(fig.\ref{telfidw}(c)). Hence we see that, overall $f^{T}$, the teleportation fidelity, fluctuates between the values $0.715$ to $0.745$, when $\omega$ and $T$ are varied together. Of course the value of $f^T$ exceeds the value of classical teleportation fidelity which is $\frac{2}{3}$ i.e. $0.67$ (approx.). When $W$ state defined in eq.(\ref{wpure}) is taken into consideration and the parties are removed from it (one at a time) to get a bipartite mixed state, it was already shown that such state has teleportation fidelity of $\frac{7}{9}$ i.e. $0.77$ (approx.) \cite{roy2010,roy2023}. Here we see that in case of prototype $W$ state, when two of the parties are exposed to event horizon of Schwarzschild black hole, and then parties are removed one at a time, it does not affect the teleportation fidelity of the resultant bipartite mixed states much although the it diminishes from the value of $\frac{7}{9}$. Therefore, it can be concluded that such states remain suitable for use as quantum teleportation channels\\\\
Next we plot the teleportation fidelity of $\rho_{W}^{wfac}$ (and $\rho_{W}^{wfab}$) against the Dilaton parameter ($D$) and monochromatic frequency ($\omega$) fixing black hole mass $M=1$.
\begin{figure}[h]
\label{telepfiddilw}
\includegraphics[width=8.7cm]{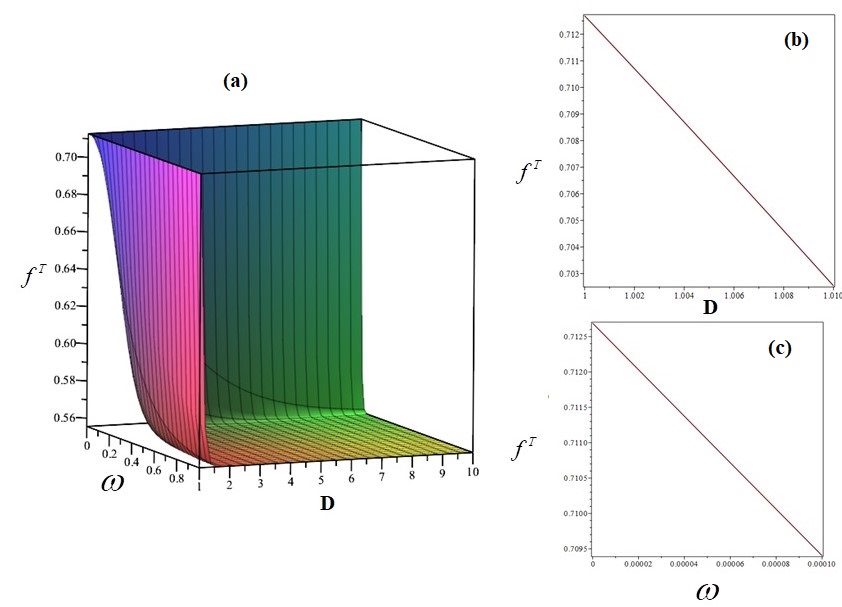}
\caption{Variation of teleportation fidelity of bipartite mixed states $\rho_{W}^{wfac}$ (or $\rho_{W}^{wfab}$) derived from tripartite prototype $W$ states in the background of Dilaton black hole. Here $D$ is varied from $1$ to $10$ and $\omega$ is varied from $0$ to $1$. Fig.(b) shows the variation of teleportation fidelity $f^T$ against $D$ and fig.(c) shows variation of $f^T$ against monochromatic frequency $\omega$.}
\label{telepfiddilw}
\end{figure}\\\\
In the above figure $\ref{telepfiddilw}(a)$, we vary monochromatic frequency ($\omega$) from $0$ to $1$ and Dilaton parameter ($D$) from $1$ to $10$, by setting unit black hole mass ($M=1$). Two dimensional overview has also been shown on the right hand side of figs.\ref{telepfiddilw}(b) and \ref{telepfiddilw}(c). Here in fig.\ref{telepfiddilw}(b), by fixing monochromatic frequency ($\omega = 1$) and subsequently varying the Dilaton paramter $D$ from $1$ to $1.01$, we see that the teleportation fidelity of $\rho_{W}^{wfac}$ (or $\rho_{W}^{wfab}$) from $0.712$ to $0.703$. On the other hand, it is observed from fig.\ref{telepfiddilw}(c), when we fix the Dilaton parameter to $D=10$ then also $f^T(\rho_{W}^{wfac})$ (or $f^T(\rho_{W}^{wfab}$) decreases from from $0.712$ to $0.709$. Therefore, in these ranges the states $\rho_{W}^{wfac}$ (or $\rho_{W}^{wfab}$) are useful as quantum teleportation channel and their teleportation fidelities exceed classical fidelity of $\frac{2}{3}$. In the context of Schwarzschild black hole, $0.715<f^T(\rho_{W}^{wfac})<0.745$ while in case of Dilaton black hole $0.709<f^T(\rho_{W}^{wfac})<0.712$. Similar is the case regarding the teleportation fidelity of $\rho_{W}^{wfab}$. \\\\
Therefore, we propose that bipartite mixed states originating from the tripartite $W$ state, even after exposure to the intense gravitational fields of Schwarzschild or Dilaton black holes, could still serve as viable quantum teleportation channels.
\subsubsection{Non-Protoype $W$ state in Black holes:}
\noindent We shall now consider the non-prototype $W$ state which we have denoted by $W_{1}$ defined in eq.(\ref{wpure1}). The parties $A$, $B$ and $C$ share qubits among themselves and the given state has tangle ($\tau = 0$). When the qubits held by parties $B$ and $C$ are exposed to event horizon of black holes and qubit held by party $A$ is in the flat region, the wave function of the state is given as
\begin{eqnarray}
\label{nonprotow}
\vert W_{1}\rangle_{wf} &=& \frac{1}{\sqrt{2}}\Big[\mu^2\vert 10000\rangle_{AB\bar{B}C\bar{C}} + \mu\nu\vert 10011\rangle_{AB\bar{B}C\bar{C}} + \mu\nu\vert 11100\rangle_{AB\bar{B}C\bar{C}} + \nu^2\vert 11111\rangle_{AB\bar{B}C\bar{C}}+\nonumber\\&& + \mu\vert 01000\rangle_{AB\bar{B}C\bar{C}} + \nu\vert 01011\rangle_{AB\bar{B}C\bar{C}} + \sqrt{2}\mu\vert 00010\rangle_{AB\bar{B}C\bar{C}} + \sqrt{2}\nu\vert 01110\rangle_{AB\bar{B}C\bar{C}}\Big.]
\end{eqnarray}
Tracing out the inaccessible regions of $B$ and $C$ i.e. $\bar{B}$ and $\bar{C}$ subsequently we get
\begin{eqnarray}
\label{denw1}
\rho_{W_{1}}^{wfabc}=\frac{1}{4}\begin{pmatrix}
0  & 0 & 0 & 0& 0 & 0 & 0 & 0\\
0  & 2\mu^2 & \sqrt{2}\mu^2 & \sqrt{2}\mu^3 & 0 & 0 & 0 & 0\\
0  & \sqrt{2}\mu^2 & \mu^2 & 0& 0 & 0 & 0 & 0\\
0  & \sqrt{2}\mu^3 & \mu^3 & \mu^4 & 0 & 0 & 0 & 0\\
0  & 0 & 0 & 0& 3\nu^2 & \mu\nu^2 & \sqrt{2}\mu\nu^2 & 0\\
0  & 0 & 0 & 0& 2\mu\nu^2 & \mu^2\nu^2 & 0 & 0\\
0  & 0 & 0 & 0& \sqrt{2}\mu\nu^2 & 0 & \mu^2\nu^2 & 0\\
0  & 0 & 0 & 0& 0 & 0 & 0 & \nu^4\\
\end{pmatrix}.
\end{eqnarray}
As before we will now trace out party $B$ from the eq. (\ref{denw1}) and thus the resultant state obtained, which is a bipartite mixed state, is shown below. 
\begin{eqnarray}
\label{rhoabipartite2w1}
\rho_{W_{1}}^{wfac}=\frac{1}{4}\begin{pmatrix}
\mu^2  & 0 & 0 & 0\\
0  & 2\mu^2+3\nu^2 & \sqrt{2}\mu^3+\sqrt{2}\mu\nu^2 & 0\\
0  & \sqrt{2}\mu^3+\sqrt{2}\mu\nu^2 & \mu^4+\mu^2\nu^2 & 0\\
0  & 0 & 0 &  \mu^2\nu^2+\nu^4 \\
\end{pmatrix},
\end{eqnarray} 
As we explained earlier, we would not trace out $A$ from the tripartite state as $A$ is in the flat region. Using eq.(\ref{concurrence}), we can calculate the concurrence of the state (\ref{rhoabipartite2w1}) and plot the result below.
\begin{figure}[h]
\label{concbipartw1}
\includegraphics[width=8.7cm]{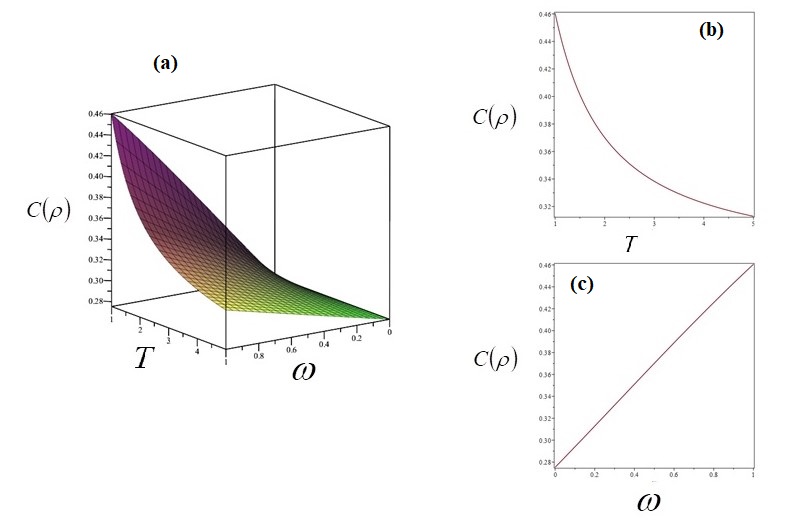}
\caption{Variation of concurrence of bipartite mixed states $\rho_{W_{1}}^{wfac}$ derived from tripartite non-prototype $W_{1}$ state in the background of Schwarzschild black hole. The Hawking temperature ($T$) is varied from $1$ to $5$ and monochromatic frequency ($\omega$) is varied from $0$ to $1$.}
\label{concbipartw1}
\end{figure}\\\\
In the fig.\ref{concbipartw1} we show the effects on the concurrence of the bipartite mixed state ($\rho_{W_{1}}^{wfac}$) when the parties $B$ and $C$ are subjected to the event-horizon of Schwarzschild black hole, though the concurrence of the state is reduced. The concurrence of the bipartite mixed state reaches is maximum value at $0.46$ when $T=1$ and $\omega = 1$. When $0\leq \omega\leq 1$, $0\leq C(\rho_{W_{1}}^{wfac})\leq 0.46$ whereas for $1\leq T\leq 5$ also $0\leq C(\rho_{W_{1}}^{wfac})\leq 0.46$. \\\\
Next we plot the concurrence of the state  $\rho_{W_{1}}^{wfac}$ when the state $W_{1}$ of eq.(\ref{nonprotow}) comes close contact with Dilaton black hole. We plot $C(\rho_{W_{1}}^{wfac})$ against the Dilaton parameter ($D$) and monochromatic frequency ($\omega$). 
\begin{figure}[h]
\label{concdilw1}
\includegraphics[width=7.5cm]{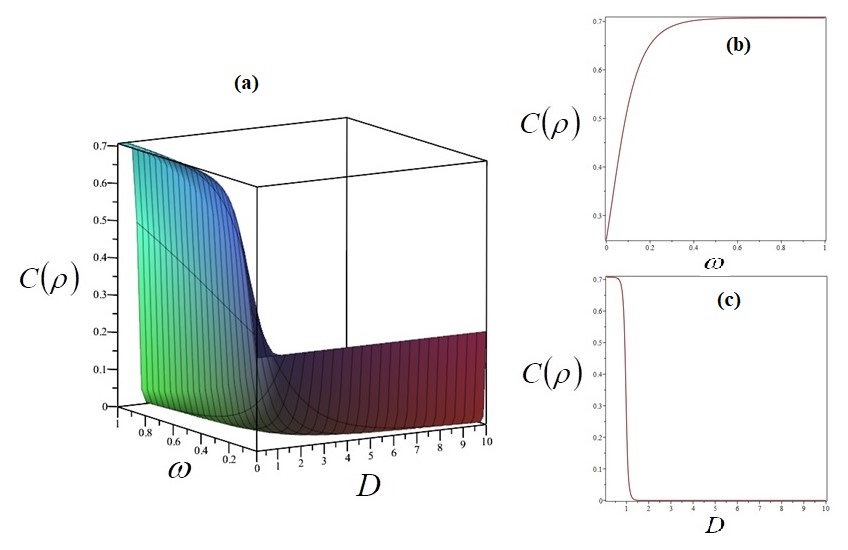}
\caption{Variation of concurrence of bipartite mixed states $\rho_{W_{1}}^{wfac}$ derived from tripartite non-prototype $W_{1}$ state in the background of Dilaton black holes. The Dilaton parameter ($D$) is varied from $0.1$ to $10$ and monochromatic frequency ($\omega$) is varied from $0$ to $1$.}
\label{concdilw1}
\end{figure}\\\\
We observe from fig.\ref{concdilw1} that when $D$ is varied from $0.1$ to $10$ then $C(\rho_{W_{1}}^{wfac})$ varies between $0$ and $0.7$. More specifically, $C(\rho_{W_{1}}^{wfac})$ starts with the value $0.7$ and then gradually starts decreasing as $D$ increases, and for $D>1.5$, it completely decays. On the other hand for $0\leq \omega \leq 1$, $0\leq C(\rho_{W_{1}}^{wfac}) \leq 0.7$.
We now check on the capacity of the bipartite mixed state $\rho_{W_{1}}^{wfac}$ (in the background of Schwarzschild black hole)  as quantum teleportation channel which we quantified by teleportation fidelity.\\\\
Using eq.(\ref{telepfid}) we calculate $f^{T}(\rho_{W_{1}}^{wfac})$ and plot this against the parameters such as monochromatic frequency ($\omega$) and the Hawking temperature ($T$). The plot is shown below. 
\begin{figure}[h]
\label{telepfidw1sch}
\includegraphics[width=8.7cm]{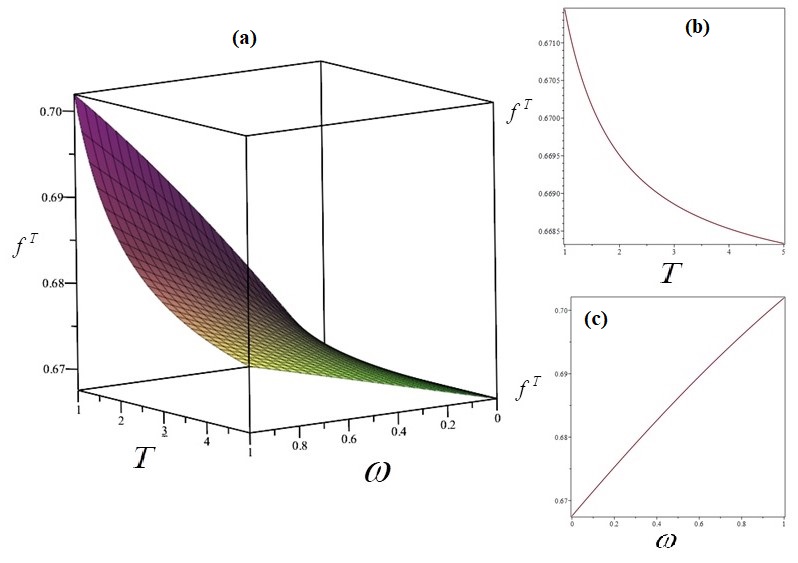}
\caption{Variation of \textcolor{blue}{teleportation fidelity} of bipartite mixed states $\rho_{W_{1}}^{wfac}$ derived from tripartite non-prototype $W_{1}$ state in the background of Schwarzschild black hole. Here $T$ is varied from $1$ to $5$ and $\omega$ is varied from $0$ to $1$, where fig.(b) shows 2D plot of teleportation fidelity with respect to Hawking temperature ($T$) and fig.(b) shows the same with respect to monochromatic frequency ($\omega$).}
\label{telepfidw1sch}
\end{figure}\\\\
The fig.\ref{telepfidw1sch} clearly shows that the as we vary Hawing temperature ($T$) from $1$ to $5$ and monochromatic frequency ($\omega$) from $0$ to $1$. the teleportation fidelity $f^{T}(\rho_{W_{1}}^{wfac})> 0.7$. Similarly we plot the teleportation fidelity $f^{T}(\rho_{W_{1}}^{wfac})$ in the system of Dilaton black hole, with respect to parameters $D$ and $\omega$.
\begin{figure}[h]
\label{telepfidw1dilt}
\includegraphics[width=8.7cm]{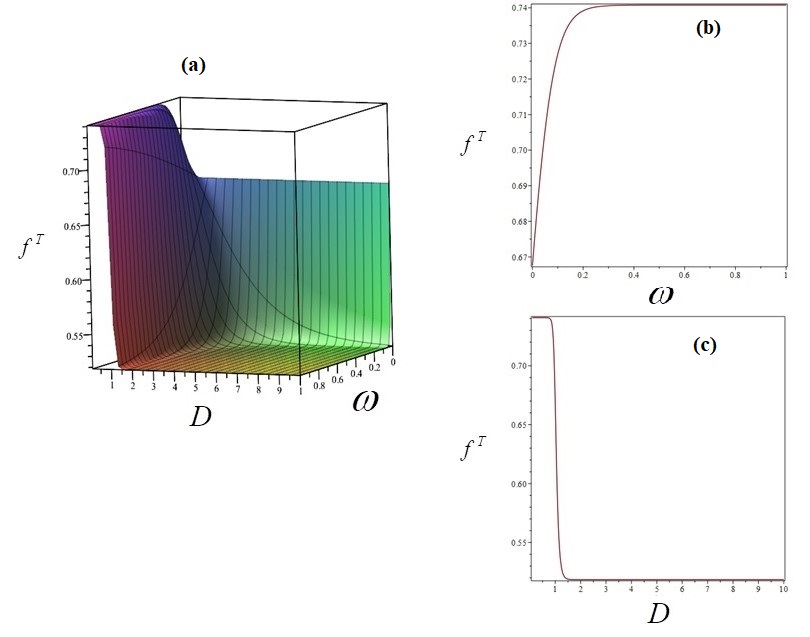}
\caption{Variation of \textcolor{blue}{teleportation fidelity} of bipartite mixed states $\rho_{W_{1}}^{wfac}$ derived from tripartite non-prototype $W_{1}$ state in the background of Dilaton black hole. Here $D$ is varied from $0.1$ to $10$ and $\omega$ is varied from $0$ to $1$, where fig.(b) shows 2D plot of teleportation fidelity with respect to monochromatic frequency ($\omega$) and fig.(c) shows the same with respect to Dilaton parameter ($D$).}
\label{telepfidw1dilt}
\end{figure}\\\\
From the fig.\ref{telepfidw1dilt}, it is clear that the teleportation fidelity of the bipartite mixed state derived from non-prototype $W$ state which was exposed to Dilaton black hole, exceeds $0.7$ depending upon the specific ranges of the Dilaton parameter ($D$) and monochromatic frequency ($\omega$).\\\\
Based on our analysis presented in figs. \ref{conc2plot},\ref{conc2plotb},\ref{telfidw},\ref{telepfiddilw} and figs.\ref{concbipartw1},\ref{concdilw1},\ref{telepfidw1sch},\ref{telepfidw1dilt}, we compare the behaviors of concurrences and fidelities of prototype $W$ state and non-prototype $W_1$ state. The non-prototype $W_1$ state shows higher maximum concurrence ($C \approx 0.46$ in Schwarzschild, $C \approx 0.70$ in Dilaton) compared to the prototype $W$ state ($C \approx 0.26$ in Schwarzschild, $C \approx 0.12$ in Dilaton), indicating stronger residual bipartite entanglement. One can also observe the prototype $W$ state maintains finite concurrence for $D < 1.16$, while the $W_1$ state extends to $D < 1.5$, showing greater resilience to the Dilaton parameter. The prototype $W$ state exhibits teleportation fidelity in the range $0.715 < f^T < 0.745$ for Schwarzschild and $0.709 < f^T < 0.712$ for Dilaton, with systematic variation against $T$ and $\omega$. The $W_1$ state maintains $f^T > 0.7$ across both backgrounds when entanglement persists, but shows sharper fidelity degradation near the critical Dilaton value. The prototype $W$ state offers more reliable performance across wider parameter ranges, making it suitable for scenarios with uncertain black hole parameters. The $W_1$ state provides higher potential fidelity but requires more controlled conditions, particularly regarding the Dilaton parameter. These differences arise from the distinct entanglement distributions in the two $W$ class states, demonstrating that the internal structure of tripartite entanglement significantly influences the teleportation capability of derived bipartite channels under gravitational effects.

\section{Conclusion:}
\noindent We have investigated the feasibility of quantum teleportation in the vicinity of black holes by analyzing bipartite mixed states derived from tripartite GHZ and W-class entangled states. In our setup, two observers (Bob and Cliff) approach the event horizon of either a Schwarzschild or a GHS Dilaton black hole, while a third (Alice) remains in flat spacetime. After tracing out one party, we evaluated the resulting bipartite state’s suitability as a teleportation channel using teleportation fidelity as the benchmark. Our results demonstrate that, despite the degradation of bipartite entanglement (quantified by concurrence) due to gravitational effects and Hawking radiation, teleportation fidelity remains above the classical threshold of $\frac{2}{3}$ for channels derived from W-class states. In contrast, GHZ-derived channels fail to support teleportation due to their lack of residual bipartite entanglement. Specifically, for Schwarzschild black holes, teleportation fidelity lies between 0.715–0.745, while for Dilaton black holes it ranges between 0.709–0.712 across the studied parameter regimes of Hawking temperature $T$, monochromatic frequency $\omega$, and Dilaton parameter $D$. This indicates that quantum teleportation remains viable near black holes provided the initial tripartite state retains useful bipartite entanglement after partial tracing. The resilience of teleportation fidelity, even as concurrence diminishes, suggests that quantum communication protocols can withstand certain relativistic gravitational effects.
While this work focused on the scenario where two parties (B and C) are near the horizon, the framework can be readily adapted to other configurations. For instance, if only Bob is near the horizon while Alice and Cliff remain in flat spacetime, the degradation would be isolated to a single party. In that case, the bipartite channel between Alice and Cliff would be expected to retain higher fidelity than the channels studied here, offering a more robust pathway for teleportation. Comparing such asymmetric configurations could provide deeper insights into how the location of gravitational degradation affects the distribution of entanglement and the viability of quantum networks in relativistic settings.
Our static analysis assumes stationary observers at fixed $r\gtrsim r_h$, capturing snapshot entanglement/fidelity without dynamical infall or LOCC signal delays. Realistically, continuous horizon approach degrades states during finite classical communication, potentially dropping $f^T<2/3$ mid protocol.
Extending our analysis to dynamical geometries with Hawking backreaction and quantum fluctuations is a key future direction. This would assess if teleportation fidelity stays above classical thresholds during evaporation and how the finite teleportation window scales with initial parameters, illuminating quantum information-gravity interplay in realistic astrophysical scenarios. Our study thus establishes a foundation for further exploration of quantum information processing in curved spacetime. Future work could extend this analysis to other tripartite or multipartite states, alternative black hole models (e.g., rotating or charged black holes), and other quantum information protocols such as quantum key distribution and dense coding. Investigating the mathematical relationship between teleportation fidelity and quantum-gravitational parameters also remains an open direction.

\vskip 0.5cm
{\bf Declaration of competing interest} The authors declare that they have no known competing financial interests or
personal relationships that could have appeared to influence the work reported in this paper.
\vskip 0.5cm
{\bf Data availability statement} All data that support the findings of this study are included within the article. No supplementary file has been added.

\end{document}